# Two Pairwise Iterative Schemes For High Dimensional Blind Source Separation

Zaid Albataineh[1] and Fathi M. Salem[2]

*Abstract*—This paper addresses the high dimensionality problem in blind source separation (BSS), where the number of sources is greater than two. Two pairwise iterative schemes are proposed to tackle this high dimensionality problem. The two pairwise schemesrealize non-parametric independent component analysis (ICA) algorithms based on a new high-performance Convex Cauchy–Schwarz Divergence (CCS-DIV). These two schemes enable fast and efficient demixing of sources in real-world high dimensional source applications. Finally, the performance superiority of the proposed schemes is demonstrated in metric-comparison with FastICA, RobustICA, convex ICA (C-ICA), and other leading existing algorithms.

## I. INTRODUCTION

Blind Signal Separation (BSS) is one of the most challenging areas in signal processing. BSS remains an important area of research and development in many domains, e.g. biomedical engineering, image processing, communication system, speech enhancement, remote sensing, etc. BSP techniques do not assume full *apriori* knowledge about the mixing environment, source signals, etc. BSS includes two major domains: Independent Component Analysis (ICA), and Multichannel Blind Deconvolution (MBD) [1-2].

In the following, we provide a focused and a brief overview. ICA is considered a key approach of BSS and unsupervised learning algorithms . ICA relagates to Principal Component Analysis (PCA) and Factor Analysis (FA) in multivariate analysis and data miningwhenthe components are in the form of Gaussian distributions [5 - 7]. However, ICA is a technique that includes higher order statistics (HOS) where, in the static mixing case, the goal is to represent a set of random variables as a linear transformation of statistically independent components.

ICA techniques are based on the assumption of non-Gaussianity and independence of the sources. Let an $M$ observation vector $x = [x_1, x_2, ... x_M]^T$ be obtained from $M$ statistically independent sources $s = [s_1, s_2, ... s_M]^T$ by the relation $x = As$, where $A$ is an $M \times M$ unknown invertible mixing matrix. The estimated (original) sources can be estimated by $y = Wx$ where $W$ is a demixing (filter) matrix. The goal in ICA is to determine a demixing matrix $W$ to estimate the source signals. ICA uses the non-Gaussianity of sources and a dependency measure to find a demixing matrix $W$. A measure, e.g., could be based on the mutual information [8 - 15], Higher Order Statistic (HOS), such as the kurtosis [5-7], orJoint Approximate Diagonalization [25-27]. In essence, the demixing matrix is obtained by optimizing such a contrast function.

Furthermore, the metrics of cumulants, likelihood function, negentropy, kurtosis, and mutual information have been developed to obtain a demixing matrix in different adaptations of ICA-based algorithms. FastICA [5], [27] was developed to maximize non-Gaussianity with relative speed and simplicity. Recently, Zarzosoand Comon [6] proposed the Robust Independent Component Analysis (R-ICA) method for betterconvergence performance.They used a truncated polynomial expansion, rather than the output marginal probability density functions, to simplify the estimation process. Moreover, in [20], the authors developed the rapid ICA algorithm which takes advantage of multi-step past information with respect to afixed-point method in order to augment the non-Guassianityamong the estimated signals. In [10–12], the authors have presented ICA methods using mutual information. They constructed a formulation by minimizing the difference between the joint entropy and the marginal entropy among the estimated sources. Moreover, the Euclidean distance divergence (ED-DIV) and the Kullback divergence (Kl-DIV) were used as the measure functions for nonnegative matrix factorization (NMF) problems in [9].

Key performance differentiators among these approaches are (i) the quality of the estimated demixed signals, and (ii) the speed of computation. In light of the advancement of computational resources, speed is nowa non-issue of most applications. However, the quality of the estimated signals is becoming of utmost importance and differentiation among the proposed methods.

[1] Electronics Engineering Department, Hijjawi Faculty for Engineering Technology, Yarmouk University, Irbid, Jordan
  Email: albatain@msu.edu,

[2] Circuits, Systems, And Neural Networks (CSANN) Laboratory
  Department of Electrical and Computer Engineering, Michigan State University, East Lansing, Michigan 48824-1226, U.S.A.
  Email: salemf@msu.edu

The present work focuses on enhancing the performance in terms of the quality of the estimated demixedsignals. To that end, wehave developed a novel contrast function for ICA based on the conjunction of a convex function into a Cauchy-Schwarz inequality-based divergence measure [23-24]. This convex Cauchy-Schwarz divergence ICA is denoted by CCS-ICA. This contrast function (CCS-ICA) is controlled by a"convexity" parameter.It also uses the Parzenwindow density modeling in order to estimatethe non-Gaussian structure of the unknown source densities. The CCS-ICA has succeeded in solving the BSS problem and it has shown a better performance than other ICA-based methods. The efficacy of the proposed method is evaluated by extensive numerical studies [28].

The paper is organized as follows. In Section II,the new convex Cauchy-Schwarz divergence measure is presented. In Section III, the new convex Cauchy-Schwarz divergence measureis specialized to the ICA domain.Section IV presents the CCS-ICA based methods. Comparative simulation results and conclusion are given in Section V and Section VI, respectively.

## II. THE CCS-DIV MEASURE

While there exists a wide range of measures, performance in terms of the quality of estimated signals, especially in audio and speech applications still in need of improvements.An improved measure should provide geometric properties for a contrast function in anticipation of a dynamic (e.g., gradient) search in the parameter space of demixing matrices. The motivation here is to introduce a simple measure and incorporate controllable curvature in order to control convergence speed to an optimal solution. To improve the divergence measure and speed up convergence to a solution, we have conjugatedconvex functions into (not merely applying it to) the Cauchy–Schwarz inequality. In this form, one takes advantage of a convexity parameter, say alpha, to control the degree of local convexity in the divergence measure and to speed up the convergence in the corresponding ICA and Non-negative Matrix Factorization (NMF) algorithms. As an example, incorporating the joint distribution ($P_J = p(z_1, z_2)$) and the marginal distributions ($Q_M = p(z_1)p(z_2)$) into the convex function, say,$f(.)$and conjugating them to the Cauchy–Schwartz inequality yields

$$|\langle f(P_J), f(Q_M) \rangle|^2 \leq \langle f(P_J), f(P_J) \rangle \cdot \langle f(Q_M), f(Q_M) \rangle$$

$$|\langle f(p(z_1, z_2)), f(p(z_1)p(z_2)) \rangle|^2 \leq \langle f(p(z_1, z_2)), f(p(z_1, z_2)) \rangle \cdot \langle f(p(z_1)p(z_2)), f(p(z_1)p(z_2)) \rangle \quad (1)$$

where$\langle \cdot , \cdot \rangle$ is an inner product; f(.) is a convex function, e.g.,

$$f(t) = \frac{4}{1-\alpha^2}\left[\frac{1-\alpha}{2} + \frac{1+\alpha}{2}t - t^{\frac{1+\alpha}{2}}\right] \text{ For} t \geq 0 \quad (2)$$

Now, based on the Cauchy–Schwartz inequality a new symmetric divergence measure is proposed, namely:

$$D_{CCS}(P_J, Q_M, \alpha) = \log \frac{\iint f^2(P_J) dz_1 dz_2 \cdot \iint f^2(Q_M) dz_1 dz_2}{[\iint f(P_J) \cdot f(Q_M) \, dz_1 dz_2]^2}$$
$$= \log \frac{\iint f^2(p(z_1, z_2)) dz_1 dz_2 \cdot \iint f^2(p(z_1) \cdot p(z_2)) dz_1 dz_2}{[\iint f(p(z_1, z_2)) \cdot f(p(z_1)p(z_2)) \, dz_1 dz_2]^2} \quad (3)$$

where, as usual,$D_{CCS}(P_J, Q_M, \alpha) \geq 0$ and equality holds if and only if $P_J = Q_M$, which means that they are independent of each other. This divergence function is then used to develop the corresponding ICA and NMF algorithms. It is notedthat the joint distribution and the product of the marginal densities in $D_{CCS}(P_J, Q_M, \alpha)$is symmetric. This symmetrical property does not hold for the Kullback–Leibler (KL) divergence (KL-DIV) [3], alpha divergence (α-DIV) [11], and f-divergence (f-DIV) [2] [13]. We anticipate that it would be desirable in the geometric structure of the search space as it would result in similar behavior from all initial conditions. Additionally, the CCS-DIV is tunable by the convexity parameter, in this case α. In contrast to the Convex divergence(C-DIV) [15] and the α-DIV [18], the range of the convexity parameter α is extendable. However, Based on l'Hopital's rule, one can derive the realization of CCS-DIV forthe case of $\alpha = 1$ and $\alpha = -1$ by finding the derivatives,with respect to $\alpha$, of the numerator and denominator for each part of $D_{CCS}(P_J, Q_M, \alpha)$. Thus, the CCS-DIV with$\alpha = 1$ and $\alpha = -1$ are respectively given by (4) and (5).

## III. THE CONVEX CAUCHY–SCHWARZ DIVERGENCE INDEPENDENT COMPONENT ANALYSIS (CCS–ICA)

### A. Non-Parametric ICA algorithms

Without loss of generality, we develop the ICA algorithm by using the CCS-DIV as a contrast function. Let us consider a simple system that is described by the vector-matrix form

$$\mathbf{x} = \mathbf{Hs} + \mathbf{v} \quad (6)$$

where $\mathbf{x} = [x_1, ..., x_M]^T$ is a mixture observation vector, $\mathbf{s} = [s_1, ..., s_M]^T$ is a source signal vector, $\mathbf{v} = [v_1, ..., v_M]^T$ is

$$D_{CCS}(P_J, Q_M, 1) = \log \frac{\left(\iint \{(p(z_1, z_2) \cdot \log(p(z_1, z_2)) - p(z_1, z_2) + 1)^2\} dz_1 dz_2\right) \cdot \left(\iint \{(p(z_1) \cdot p(z_2) \cdot \log(p(z_1) \cdot p(z_2)) - p(z_1) \cdot p(z_2) + 1)^2\} dz_1 dz_2\right)}{[\iint \{(p(z_1, z_2) \cdot \log(p(z_1, z_2)) - p(z_1, z_2) + 1) \cdot (p(z_1) \cdot p(z_2) \cdot \log(p(z_1) \cdot p(z_2)) - p(z_1) \cdot p(z_2) + 1)\} dz_1 dz_2]^2} \quad (4)$$

$$D_{CCS}(P_J, Q_M, -1) = \log \frac{\left(\iint \{(\log(p(z_1, z_2)) - p(z_1, z_2) + 1)^2\} dz_1 dz_2\right) \cdot \left(\iint \{(\log(p(z_1) \cdot p(z_2)) - p(z_1) \cdot p(z_2) + 1)^2\} dz_1 dz_2\right)}{[\iint \{(\log(p(z_1, z_2)) - p(z_1, z_2) + 1) \cdot (\log(p(z_1) \cdot p(z_2)) - p(z_1) \cdot p(z_2) + 1)\} dz_1 dz_2]^2} \quad (5)$$

an additive (Gaussian) noise vector, and $\mathbf{H}$ is an unknown full rank M × M mixing matrix, where M is the number of source signals. To obtain a good estimate, $\mathbf{y} = \mathbf{Wx}$ of the source signals $\mathbf{s}$, the contrast function CCS-DIV should be minimized with respect to the demixing filter matrix $\mathbf{W}$. Thus, the components of $\mathbf{y}$ become least dependent when this demixing matrix $\mathbf{W}$ becomes a scaled permutation of $\mathbf{H}^{-1}$. Following the standard ICA procedure, the estimated source $\mathbf{y}$ can be carried out in two steps: 1) the original data $\mathbf{x}$ should be preprocessed by removing the mean, i.e. one assumes $\{E[\mathbf{x}] = 0\}$ and also by a (pre-)whitening matrix $\{\mathbf{V} = \mathbf{\Lambda}^{-1/2}\mathbf{E}^T\}$, where the matrix $\mathbf{E}$ represents the eigenvectors and $\mathbf{\Lambda}$ the eigenvalues matrices of the autocorrelation, namely an estimate of $\{\mathbf{R_{xx}} = E[\mathbf{xx}^T]\}$. Consequently, the whitened data vector $\{\mathbf{x}_t = \mathbf{Vx}\}$ would have zero-mean and its covariance equal to the identity matrix, i.e., $\{\mathbf{R}_{\mathbf{x}_t\mathbf{x}_t} = \mathbf{I_M}\}$. The demixing matrix can be iteratively computed by, e.g., the gradient descent algorithm [2]:

$$\mathbf{W}(k+1) = \mathbf{W}(k) - \gamma \frac{\partial D_{CCS}(X,\mathbf{W}(k))}{\partial \mathbf{W}(k)} \quad (7)$$

where $k$ represents the iteration index and $\gamma$ is a step size or a learning rate. Therefore, the updated term in the gradient descent is composed of the differentials of the CCS-DIV with respect to each element $w_{ml}$ of the M × M demixing matrix $\mathbf{W}$. The differentials $\frac{\partial D_{CCS}(X,\mathbf{W}(k))}{\partial w_{ml}(k)}$, $1 \leq m, l \leq M$ are calculated using a probability model and the CCS-DIV measure as in [7] and [14]. However, the update procedure may be stopped, e.g., when the absolute increment of the CCS-DIV measure meets a predefined threshold value. During iterations, one should make the normalization step $\mathbf{w}_m = \mathbf{w}_m / \|\mathbf{w}_m\|$ for each row of $\mathbf{W}$, where $\|.\|$ denotes a norm. Please refer to *Algorithm 1* for the delineated algorithm based on gradient descent.

In deriving the CCS–ICA algorithm, based on the proposed CCS-DIV measure $D_{CCS}(\mathbf{P}_J, \mathbf{Q}_M, \alpha)$, usually, the vector $\mathbf{P}_J$ corresponds to the probability of the observed data $\left(p(\mathbf{y}_t) = p(\mathbf{Wx}_t) = \frac{p(\mathbf{x}_t)}{|\det(\mathbf{W})|}\right)$ and vector $\mathbf{Q}_M$ corresponds to the probability of the estimated or expected data $(\prod_1^M p(y_{mt}) = \prod_1^M p(\mathbf{w}_m \mathbf{x}_t))$. Here, the CCS–ICA algorithm is detailed as follows. Let the demixed signals $\mathbf{y}_t = \mathbf{Wx}_t$ with its $mth$ component denoted as $y_{mt} = \mathbf{w}_m \mathbf{x}_t$. Then, $\mathbf{P}_J = p(\mathbf{y}_t) = p(\mathbf{Wx}_t)$ and $Q_M = \prod_1^M p(y_{mt}) = \prod_1^M p(\mathbf{w}_m \mathbf{x}_t)$. Thus, the CCS-DIV as the contrast function with the built-in convexity parameter $\alpha$, is

$$\begin{aligned} &D_{CCS}(\mathbf{P}_J, Q_M, \alpha) \\ &= \log \frac{\iint f^2(\mathbf{P}_J) dy_1 \dots dy_M \cdot \iint f^2(Q_M) dy_1 \dots dy_M}{[\iint f(\mathbf{P}_J) \cdot f(Q_M) dy_1 \dots dy_M]^2} \\ &= \log \frac{\iint f^2(p(\mathbf{Wx}_t)) dy_1 \dots dy_M \cdot \iint f^2(\prod_1^M p(y_{mt})) dy_1 \dots dy_M}{[\iint f(p(\mathbf{Wx}_t)) \cdot f(\prod_1^M p(y_{mt})) dy_1 \dots dy_M]^2} \end{aligned}$$
(8)

For any convex function, we use the Lebesgue measure to approximate the integral with respect to the joint distribution of $y_t = \{y_1, y_2, \dots, y_M\}$. The contrast function thus becomes

$$\begin{aligned} D_{CCS}(\mathbf{P}_J, Q_M, \alpha) &= \log \frac{\sum_1^T f^2(p(\mathbf{Wx}_t)) \cdot \sum_1^T f^2(\prod_1^M p(y_{mt}))}{[\sum_1^T f(p(\mathbf{Wx}_t)) \cdot f(\prod_1^M p(y_{mt}))]^2} \\ &= \log \frac{\sum_1^T f^2(p(\mathbf{Wx}_t)) \cdot \sum_1^T f^2(\prod_1^M (p(\mathbf{w}_{mt}\mathbf{x}_t)))}{[\sum_1^T f(p(\mathbf{Wx}_t)) \cdot f(\prod_1^M (p(\mathbf{w}_{mt}\mathbf{x}_t)))]^2} \end{aligned}$$
(9)

The adaptive CCS–ICA algorithms are carried out by using the derivatives of the proposed divergence, i.e., $\left(\frac{\partial D_{CCS}(\mathbf{P}_J, Q_M, \alpha)}{\partial w_{ml}}\right)$ as derived in Appendix A. Note that in Appendix A, the derivative of the determinant demixing matrix $(\det(\mathbf{W}))$ with respect to the element $(w_{ml})$ equals the cofactor of entry $(m, l)$ in the calculation of the determinant of $\mathbf{W}$, which we denote as $\left(\frac{\partial \det(\mathbf{W})}{\partial w_{ml}} = W_{ml}\right)$. Also the joint distribution of the output is determined by $p(\mathbf{y}_t) = \frac{p(\mathbf{x}_t)}{|\det(\mathbf{W})|}$.

For simplicity, we can write $D_{CCS}(\mathbf{P}_J, Q_M, \alpha)$ as a function of three variables.

$$D_{CCS}(\mathbf{P}_J, Q_M, \alpha) = \log \frac{V_1 \cdot V_2}{(V_3)^2} \quad (10)$$

Then,

$$\frac{\partial D_{CCS}(\mathbf{P}_J, Q_M, \alpha)}{\partial w_{ml}} = \frac{V_1' V_2 + V_1 V_2' - 2V_1 V_2 V_3'}{V_1 V_2 V_3} \quad (11)$$

where

$$V_1 = \sum_{t=1}^T f^2(\mathbf{P}_J), \quad V_1' = \sum_{t=1}^T 2f(\mathbf{P}_J)f'(\mathbf{P}_J)\mathbf{P}_J'$$

$$V_2 = \sum_{t=1}^T f^2(Q_M), \quad V_2' = \sum_{t=1}^T 2f(Q_M)f'(Q_M)Q_M'$$

$$V_3 = \sum_{t=1}^T f(\mathbf{P}_J)f(Q_M),$$

$$V_3' = \sum_{t=1}^T f'(\mathbf{P}_J)f(Q_M)\mathbf{P}_J' + \sum_{t=1}^T f(\mathbf{P}_J)f'(Q_M)Q_M'$$

$$\mathbf{P}_J = p(\mathbf{Wx}_t) \text{ and } Q_M = \prod_{m=1}^M p(\mathbf{w}_m \mathbf{x}_t)$$

$$\mathbf{P}_J' = \frac{\partial \mathbf{P}_J}{\partial w_{ml}} = -\frac{p(\mathbf{x}_t)}{|\det(\mathbf{W})|^2} \cdot \frac{\partial \det(\mathbf{W})}{\partial w_{ml}} \cdot \text{sign}(\det(\mathbf{W})),$$

where $\frac{\partial \det(\mathbf{W})}{\partial w_{ml}} = W_{ml}$.

$$Q_M' = \frac{\partial Q_M}{\partial w_{ml}} = \left[\prod_{j=m}^M p(\mathbf{w}_j \mathbf{x}_t)\right] \frac{\partial p(\mathbf{w}_n \mathbf{x}_t)}{\partial (\mathbf{w}_n \mathbf{x}_t)} \cdot x_l.$$

Where $x_l$ denotes the $lth$ entry of $\mathbf{x}_t$.

In general, the estimation accuracy of a demixing matrix in the ICA algorithm is limited by the lack of knowledge of

## Algorithm 1: the ICA Based algorithm using the gradient descent

**Input:** $(M \times T)$ matrix of realizations $X$, Initial demixing matrix $W = I_M$, Max. number of iterations Itr, Step Size $\gamma$ i.e. $\gamma = 0.3$, alpha $\alpha$ i.e. $\alpha = -0.99999$

**Perform Pre-Whitening**
$$\{X = V * X = \Lambda^{\wedge}(-1/2) \, E^{\wedge}T \, X\},$$

**For loop:** for each I Iteration do
  **For loop:** for each $t = 1, \dots, T$
  Evaluate the proposed contrast function and its derivative
$$\left(\partial D_{CCS}(P_J, Q_M, \alpha) / \partial w_{ml}\right)$$
  **End For**
  Update demixing matrix $W$
$$W = W - \gamma \frac{\partial D_{CCS}(P_J, Q_M, \alpha)}{\partial W}$$
  Check Convergence
  $\|\Delta D_c\| \leq \epsilon$ i.e. $\epsilon = 10^{-4}$
**End For**

**Output:** Demixing Matrix $W$, estimated signals $y$

---

the accurate source probability densities. However, non-parametric density estimate is used in [7], [15], by applying the effective Parzen window estimation. One of the attributes of the Parzen window is that it must integrate to one. Furthermore, it exhibits a distribution shape that is data-driven and is flexibly formed based on its chosen Kernel functions.. Thus, one can estimate the density function $p(y)$ of the process generating the $M$-dimensional sample $y_1, y_2 \dots y_M$ due to the Parzen Window estimator. For all these reasons, a non-parametric CCS–ICA algorithm is also presented by minimizing the CCS-DIV to generate the demixed signals $y = [y_1, y_2, \dots, y_M]^T$. Here, the demixed signals are described by the following univariate and multivariate distribution estimates [18],

$$p(y_m) = \frac{1}{Th} \sum_{t=1}^{T} \vartheta\left(\frac{y_m - y_{mt}}{h}\right) \quad (12)$$

$$p(y) = \frac{1}{Th^M} \sum_{t=1}^{T} \varphi\left(\frac{y - y_t}{h}\right) \quad (13)$$

where the univariate Gaussian Kernel is
$$\vartheta(u) = (2\pi)^{-\frac{1}{2}} e^{-\frac{u^2}{2}}$$

and the multivariate Gaussian Kernel is
$$\varphi(u) = (2\pi)^{-\frac{N}{2}} e^{\frac{-1}{2} u^T u}.$$

The Gaussian kernel(s), used in the non-parametric ICA, are smooth functions. We note that the performance of a learning algorithm based on the non-parametric ICA is better than the performance of a learning algorithm based on parametric ICA. By substituting (12) and (13) with $y_t = W x_t$ and $y_{mt} = w_m x_t$ into (9), the nonparametric CCS-DIV becomes

$$P_J = p(y_t) = p(W x_t) = \frac{1}{Th^M} \sum_{t=1}^{T} \varphi\left(\frac{W(x_t - x_i)}{h}\right)$$

Or
$$P_J = p(y_t) = \frac{p(x_t)}{|\det(W)|}$$

$$Q_M = \prod_{1}^{M} p(y_{mt}) = \prod_{1}^{M} p(w_m x_t)$$

$$= \prod_{1}^{M} \frac{1}{Th} \sum_{i=1}^{T} \vartheta\left(\frac{w_m(x_t - x_i)}{h}\right)$$

$$D_{CCS}(P_J, Q_M, \alpha) = \log \frac{\sum_{t=1}^{T} f^2(P_J) \cdot \sum_{t=1}^{T} f^2(Q_M)}{[\sum_{t=1}^{T} f(P_J) \cdot f(Q_M)]^2} \quad (14)$$

However, there are two common methods to minimize this divergence function: one is based on the gradient descent approach and the other is based on an exhaustive search such as the Jacobi method. We have presented the derivation of the proposed algorithm in *Appendix A* in order to use it in the non-parametric gradient descent ICA algorithm, see ***Algorithm 1***.

### B. Scenario of three source signals and more

Generally Speaking, the non-parametric ICA algorithm suffers from insufficient data and high computation in a high dimensional space, especially when estimating the joint distribution. However, in several previous reports in the literature, e.g., [8], [15], the authors suggest applying the pairwise iterative schemes to tackle the high dimensional data problem for non-parametric ICA algorithm(s). However, there are no results indicating how the performance would hold up with the pairwise scheme, especially in terms of computational complexity and in terms of the accuracy of the non-parametric ICA algorithm.

In this work, we present two effective pairwise ICA algorithms: one is based on the gradient descent and the other is based on the Jacobi optimization [4].

Without loss of generality, one can represent the demixing matrix $W$ as a series of rotational matrices in terms of an unknown angle $\theta_{ij} \in [-\pi/4, \pi/4]$ between each pair $(i, j)$ of the observed signals. Specifically, define the pairwise rotation matrix

$$W(\theta_{ij}) = \begin{bmatrix} \cos \theta_{ij} & -\sin \theta_{ij} \\ \sin \theta_{ij} & \cos \theta_{ij} \end{bmatrix} \quad (15)$$

The idea is to make each pair of the estimated (marginal) output as "independent" as possible (i.e., minimize dependency). It was proven and pointed out by Comon in [27] that the mutual independence between the M whitened observed signals can be attained by maximizing the independence between each pair of them. To that end, we present two algorithms to solve the high dimensional problem in the non-parametric scheme. First, we adopt the non-parametric algorithm based on the gradient descent into the pairwise iterative scheme of ***Algorithm 2***.

Second, we propose a CCS-ICA algorithm based on Jacobi pairwise scheme in Algorithm 3. This algorithm is based on finding the rotation matrix in (15) that attains the minima of CCS-DIV. Thus, we set up the resolution of thetas such that $\theta_{ij} \in [-\frac{pi}{4} : \theta_g : \frac{pi}{4}]$, where $\theta_g$ is the grid search, for instance $\theta_g = \frac{pi}{64}$. Then for each pair (i, j) of the observation data in the range, we find the demixing matrix $W_2$, which attains the minimum of the CCS-DIV. Please refer to **Algorithm 3** for more details.

---

**Algorithm 2: The** *ICA Based on pairwise gradient decent scheme*

---

**Input:** $(M \times T)$ matrix of realizations $X$, Initial demixing matrix $W = I_M$, number of iterations $Itr$, Step Size $\gamma$ e.g., $\gamma = 0.3$, $\alpha$, e.g., $\alpha = -0.99999$
**For** $itr = 1 \ldots itrmax$
 **Perform Pre-Whitening**
  $\{X = V * X = \Lambda^{((-1)/2)} E^T X\}$,
  **For loop:** for each $i = 1 \ldots M - 1$
   **For loop:** for each $j = i + 1 \ldots M$
    Initial demixing matrix $W_2 = I_2$
     **While:** while (true)
      Find $W_2$ e.g., using **Algorithm 1** for each pairs of $X$ ;
   **End While**
     Initial rotational matrix
  $R = I_M$,
     Update rotational matrix
      $R([i\ j], [i\ j]) = W_2$
     Update Demixing matrix
      $W = R * W$
   **End For** $j$
  **End For** $i$
 **End For** $itr$
**Output:** Demixing matrix $W = W * V$ and demixed sources in $X = W * X$

---

## C. Computational Complexity

Given $T$ realizations of $M$ observation signals, the computational complexity of the proposed algorithms rely on $T$ and the number of observation signals $M$, and is approximately given by $O\left(\frac{M(M-1)}{2} T^2\right)$. The computational complexity has been a measure of merit for ICA algorithms. With the advent of Graphics Processing Units (GPUs) (see Nvidia.com, e.g.), and more powerful computing platforms, however performance accuracy holds more merit. In our comparison among the ICA algorithms, we employ several metrics including computational time and accuracy. We also employ adaptive sampling techniques that improves the performance in terms of both metrics (accuracy and computational load). The presented technique samples the signal into small time blocks in order to evaluate the integration of the proposed divergence and reduce the computational complexity. Thus, we have introduced a sampling factor $T_s$ to evaluate the proposed divergence at each $T_s$ instance. Therefore, the computational complexity of the proposed algorithm is reduced by the square of the sample factor $T_s$ to be less than $O\left(\frac{M(M-1)}{2}\left(\frac{T}{T_s}\right)^2\right)$. Namely, we quantize the specific area of integration of the proposed divergence into equal $\left(\frac{T}{T_s}\right)$ segments to evaluate the proposed divergence.

---

**Algorithm 3: The** *ICA Based on pairwise Jacobi scheme*

---

**Input:** $(M \times T)$ matrix of realization $X$, Initial demixing matrix $W = I_M$, number of iterations $Itr$, Step Size $\gamma$, e.g., $\gamma = 0.3$, $\alpha$ e.g., $\alpha = -0.99999$
**Perform Pre-Whitening**
  $\{X = V * X = \Lambda^{((-1)/2)} E^T X\}$,
**While (True)**
 **For loop:** for each $i = 1 \ldots M - 1$
  **For loop:** for each $j = i + 1 \ldots M$
   **If** $CM([i\ j], [i\ j]) == 0$
     **Continue;**
   **end**
   **For loop:** For each $\theta_1 = -\frac{pi}{4} : \frac{pi}{64} : \frac{pi}{4}$
    $W_2 = \begin{bmatrix} \cos\theta_1 & -\sin\theta_1 \\ \sin\theta_1 & \cos\theta_1 \end{bmatrix}$
   Evaluate
   $D_c(X([i\ j],:), W_2 * X([i\ j],:), \alpha)$ For all $t = 1, \ldots, T$.
   **End For**
    Find
    $W_2 = \min_{W_2} D_c(X(i:j,:), W_2 * X, \alpha)$
   Initial rotational matrix
   $R = I_M$,
   Update rotational matrix
    $R([i\ j], [i\ j]) = W_2$
   Update Demixing matrix
    $W = R * W$
   Update Convergence matrix
    $CM([i\ j], [i\ j]) = theta * \frac{180}{pi}$
 **End For**
 **End For**
**End while loop** If $sum(CM) <= 1$
  **Output:** Demixing matrix
  $W = W * V$ and estimated Sources in $X = W * X$

---

### IV. COMPARATIVE SIMULATION STUDY

#### A. Performance evaluation of the proposed CCS-ICA algorithms versus the existing ICA-based algorithms

In this section, Monte Carlo Simulations are carried out. It is assumed that the number of sources is equal to the number of observations "i.e., sensors". All algorithms have used the same whitening method. The simulations have been carried out using the MATLAB software on an Intel Core i5 CPU 2.4-GHz processor and 4G MB RAM. Each entry in the forthcoming tables corresponds to the average of corresponding trial "independent Monte Carlo" runs in which the mixing matrix is randomly chosen.

Firstly, we start with the 2x2 mixture matrix case as a baseline for verifying the performance of the presented algorithms and thoroughly studying the impact of various classes of source signals, namely, uniform distributions, Laplacian distributions, Rayleigh distributions and log-normal distributions, on the performance of the proposed algorithm.

We compare the performance of the ICA algorithms based on the CCS-DIV, CS-DIV, E-DIV, KL-DIV, and C-DIV with $\alpha = 1$ and $\alpha = -1$ for the *Algorithm 3* scheme. We also compare it with other benchmark algorithms such as FastICA[3] [5], RobustICA[4] [6], JADE[5] [25-26] and RapidICA[6] [20]. For these methods, the default setting parameters are used according to their toolboxes and their publications.

---

[3] http://www.cis.hut.fi/projects/ica/fastica/code/dlcode.html
[4] http://www.i3s.unice.fr/~zarzoso/robustica.html
[5] http://www.tsi.enst.fr/icacentral/algos.html
[6] http://dx.doi.org/10.4236/jsip.2012.33037

During the comparison, we use the bandwidth as a function of sample size, namely, $h = 1.06T^{\frac{-1}{5}}$ [13-15]. The demixing matrix has been initialized as an identity i.e., $W = I_M$ for all algorithms. Note that CCS2 and CCS3 represent **Algorithm 2** and **Algorithm 3**, respectively. In addition, the minus and plus signs represent $\alpha = 1$ and $\alpha = -1$ cases, respectively. **Table I**, **Table II** and **Table III** summarize the performance of the proposed non-parametric ICA algorithms "CCS2 and CCS3" against other aforementioned algorithms. In this task, our goal is to separate mixtures of two sub-Gaussians, two sup-Gaussians, and both sub and sup- Gaussian signals. Specifically, we use the following distributions:

For the sub-Gaussian distribution, we use (i) the uniform distribution

$$p(s_1) = \begin{cases} \frac{1}{2\tau_1} s_1 \ in\ (-\tau_1, \tau_1) \\ 0\ \text{Otherwise} \end{cases} \quad (16)$$

and (ii) the Rayleigh distribution, i.e.,

$$p(s_2) = s_2 \exp\left[-\frac{s_2^2}{2}\right] \quad (17)$$

For the super-Gaussian distribution, we use (i) the Laplacian distribution

$$p(s_3) = \frac{1}{2\tau_2} \exp\left[-\frac{|s_3|}{\tau_2}\right] \quad (18)$$

and (ii) the log-normal distribution, i.e.,

$$p(s_4) = \exp\left[-\frac{(\log s_4)^2}{2}\right] \quad (19)$$

Also, data samples, $T = 1000$, are selected and randomly generated by using $\tau_1 = 3$ and $\tau_2 = 1$ in the above. Kurtoses for all aforementioned signals are $-1.2, 2.99, -0.7224,$ and $8.4559$ respectively, and they are evaluated using $\text{Kurt}(s) = E[s^4]/(E[s^2])^2 - 3$. Furthermore, A different, randomly generated mixing matrices are used to generate the mixtures.

One can observe several patterns from **Table I, II and III**. The presented algorithms based on the proposed measure show the best performance in terms of accuracy (in most cases) and repeatability (in terms of variance). The proposed algorithm CCS3 exhibits comparable behavior in terms of speed and stability with KL and ED. Clearly, the proposed divergence improves on the CS-DIV in terms of repeatability and performance. Notably, most the presented divergences struggle to separate the Rayleigh distributions $(s_2, s_2)$ (including the KL-DIV) except the proposed divergence and C-DIVs. Moreover, **Table III** shows the variance of the performance of the proposed algorithm outperforms the CS-DIV and renders the divergence more robust against variation in parameters. It is also worthwhile to represent the average performance of each method in Fig. 1.

Also, it is noted that the non-parametric methods result in better performance and repeatability than methods such as JADE, FastICA and other algorithms. Nevertheless JADE performs better than each of FastICA, RobustICA and Rapid ICA in terms of accuracy in some cases. However, in terms of speed, we find that these later algorithms outperform the JADE algorithm, especially rapid ICA and Robust ICA.

Secondly, An extensive analytical study is carried out to evaluate and show the performance of the proposed algorithm and other algorithms for the high dimensional case (i.e., mixture of more than 2 sources).To that end, we form five groups based on the number of source signals in the mixtures, namely, **2, 4, 8, 16** and **20**. Then, we implement ourproposed algorithms and other hosted algorithms on each corresponding group using various sample size to show the impact of the sample size on their performance.

*Table I: The performance of the ICA algorithm based on the proposed divergence and other widely used ICA algorithms in terms of the Amari error (multiplied by 100). Each entry averages over the corresponding number of trials. Observation mixtures consist of two source signals that follow the same distribution as denoted in the corresponding example.*

| Source | Samples | Trials | FastICA | JADE | RobustICA | Rapid ICA | IK-DIV | CS-DIV | KL-ICA | ED-DIV | C-DIV+ | C-DIV- | CCS-DIV1+ | CCS-DIV1- | CCS-DIV2+ | CCS-DIV2- |
|---|---|---|---|---|---|---|---|---|---|---|---|---|---|---|---|---|
| $s_1, s_1$ | 1000 | 100 | 6.16 | 4.77 | 5.27 | 5.07 | 3.32 | 2.66 | 1.75 | 2.04 | 2.17 | 2.36 | 2.25 | 2.40 | 1.86 | 1.71 |
| $s_2, s_2$ | 1000 | 100 | 22.34 | 18.51 | 28.29 | 20.26 | 6.78 | 7.39 | 8.13 | 5.12 | 5.38 | 3.83 | 8.92 | 5.80 | 3.55 | 2.99 |
| $s_3, s_3$ | 1000 | 100 | 2.45 | 2.10 | 2.24 | 2.14 | 2.31 | 2.21 | 2.31 | 2.31 | 2.65 | 2.50 | 2.19 | 1.94 | 1.84 | 1.84 |
| $s_4, s_4$ | 1000 | 100 | 3.34 | 3.03 | 3.13 | 3.29 | 1.93 | 2.02 | 1.93 | 1.71 | 2.04 | 1.90 | 1.97 | 1.93 | 1.82 | 1.76 |
| $s_1, s_3$ | 1000 | 100 | 5.11 | 4.53 | 5.39 | 5.17 | 2.44 | 2.07 | 2.06 | 2.24 | 2.56 | 2.10 | 2.50 | 2.33 | 2.21 | 1.97 |

*Table II: The computational load, in seconds, of the ICA algorithm based on the proposed divergence and other widely used ICA algorithms, each entry averages over the corresponding number of trials. Observation mixtures consists of two source signals that follow the same distribution as denoted in the corresponding example.*

| Source | Samples | Trials | FastICA | JADE | RobustICA | Rapid ICA | IK-DIV | CS-DIV | KL-ICA | ED-DIV | C-DIV+ | C-DIV- | CCS-DIV2+ | CCS-DIV2- | CCS-DIV3+ | CCS-DIV3- |
|---|---|---|---|---|---|---|---|---|---|---|---|---|---|---|---|---|
| $s_1, s_1$ | 1000 | 100 | 0.0 | 0.0 | 0.0 | 0.0 | 20.1 | 22.1 | 19.5 | 20.1 | 24.1 | 24.1 | 22.2 | 22.2 | 19.3 | 19.3 |
| $s_2, s_2$ | 1000 | 100 | 0.0 | 0.1 | 0.0 | 0.0 | 20.1 | 21.3 | 19.2 | 20.2 | 23.3 | 23.3 | 19.1 | 19.1 | 21.2 | 21.2 |
| $s_3, s_3$ | 1000 | 100 | 0.0 | 0.0 | 0.0 | 0.0 | 19.1 | 20.7 | 19.1 | 22.1 | 25.1 | 25.1 | 18.1 | 18.1 | 20.2 | 20.2 |
| $s_4, s_4$ | 1000 | 100 | 0.0 | 0.1 | 0.0 | 0.0 | 20.4 | 24.3 | 19 | 23.1 | 24.1 | 24.1 | 19.1 | 19.1 | 19.2 | 19.2 |
| $s_1, s_3$ | 1000 | 100 | 0.0 | 0.0 | 0.0 | 0.0 | 20.2 | 20.1 | 20.1 | 22.1 | 21.4 | 21.4 | 18.1 | 18.1 | 19.2 | 19.2 |

*Table III: The corresponding variance of the performance.*

| Source | Samples | Trials | FastICA | JADE | RobustICA | Rapid ICA | IK-DIV | CS-DIV | KL-ICA | ED-DIV | C-DIV+ | C-DIV- | CCS-DIV2+ | CCS-DIV2- | CCS-DIV3+ | CCS-DIV3 |
|---|---|---|---|---|---|---|---|---|---|---|---|---|---|---|---|---|
| $s_1, s_1$ | 1000 | 100 | 11.02 | 12.07 | 38.05 | 11.74 | 3.76 | 2.58 | 0.81 | 1.15 | 1.39 | 0.98 | 1.53 | 1.91 | 0.72 | 0.78 |
| $s_2, s_2$ | 1000 | 100 | 102.53 | 211.75 | 332.76 | 95.06 | 16.43 | 37.71 | 8.87 | 8.87 | 6.76 | 3.92 | 28.35 | 10.63 | 6.19 | 2.62 |
| $s_3, s_3$ | 1000 | 100 | 1.11 | 1.80 | 1.71 | 1.27 | 1.63 | 1.54 | 1.66 | 1.66 | 1.60 | 1.35 | 2.39 | 1.93 | 1.17 | 0.86 |
| $s_4, s_4$ | 1000 | 100 | 18.47 | 15.34 | 17.44 | 14.64 | 1.51 | 1.52 | 0.95 | 0.78 | 1.07 | 1.08 | 1.19 | 1.51 | 0.83 | 0.83 |
| $s_1, s_3$ | 1000 | 100 | 13.91 | 12.88 | 13.90 | 14.16 | 2.14 | 2.75 | 1.63 | 1.72 | 2.25 | 1.16 | 2.04 | 1.92 | 1.37 | 1.02 |

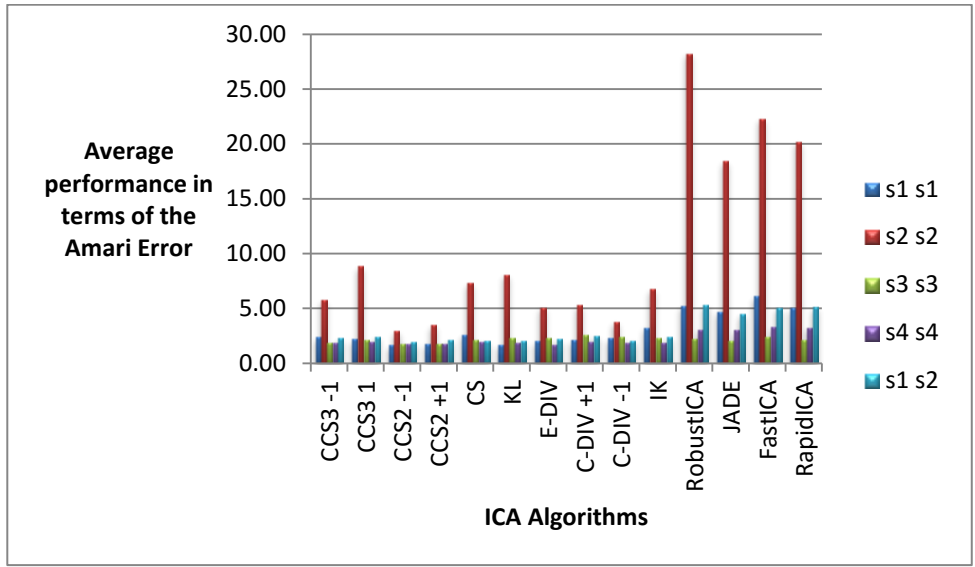

Fig. 1 Average performance of the ICA algorithms in terms of the Amari Error

A new set of randomly generated source signals (refer to *Table IV*) and mixing matrices are generated for the next comparative case study.*Table V* summarizes the performance of the aforementioned algorithms in a more complex separation process for mixtures of multiple sources. We use the label dimension to identify the number of signal sources in Table V. In a nutshell, *TableV* summarizes the performance of each algorithm in terms of the standard Ammarierror metric (multiplied × 100), see [2], [13]. All results have been averaged over a number of independent Monte Carlo runs. Several pattrens can be observed from **Table V**. First, the non-parametric ICA algorithms attain the best performance in terms of accuracy. However, in terms of speed, RapidICA, FastICA, RobustICA and JADE perform better. Second, the non-parametric ICA based on the proposed divergence (CCS3)provides the best performance in terms of accuracy (in most cases). Third, the performance of the new algorithms perform more consistently and exhibitsperformance improvement as the sample size increases. Lastly, in terms of speed, RapidICA, FastICA, RobustICA and JADE perform better. Thus, these latter algorithms could be chosento initializethe process of the proposed algorithms in order to reduce the overall computational load.

Since the comparison between the ICA algorithms has relied on two criteria, namely, accuracy and computational load, a tradeoff between these two criteria has always been assessed for eachtargeted application. We also note that with the advent of powerful computing platforms including Graphics Processing Units (GPUs), computational load/speed becomes less of a factor, and thetrue metric becomes accuracy or quality of estimates.

Furhtermore, *Table VI*summarizes the performance of CCS-ICA (see **Algorithm 3**) based on the different values of $T_s$ (1, 10, 100, 1000), which correspond to the length of a sample frame of data.Finally, *Table VII* shows the various relative computational load of the algorithms in seconds (using the same intel i5 CPU processor of a portable PC). Based on Table VI, one observes that the accuracy of the proposed algorithm degrades as the sample time frame, $T_s$,is increased. However, the speed of the proposed algorithm speeds up as the sample time $T_s$ is increased. It is observed that the CCS-ICA-- **Algorithm 3 scheme at** $T_s = 100$ attains the best performance in terms of both metrics (accuracy and computational complexity). More comparative resultson theproposed non-parametric CCS-ICA algorithmare relegated to http://www.egr.msu.edu/bsr/.

## I. CONCLUSION

Two schemes of pairwise non-parametric ICA algorithms are developed to tackle the dimensional curse in BSS. Several simulations are carried out to show the improved performance of the proposed schemes. Furthermore, the paper provides a comparative Monte-Carlometric performance with a host of leading ICA algorithms. We have also introduced sampling factor $T_s$to evaluate the CCS-DIVfor several $T_s$values.The computational complexity of the non-parametric ICA algorithm is reduced in proportion to the sample factor $T_s$.

## APPENDIX A

### CONVEX CAUCHY–SCHWARZ DIVERGENCE AND ITS DERIVATIVE

Assume the demixed signals $y_t = Wx_t$ with its*mth*component denoted as $y_{mt} = w_m x_t$.Then, $P_J =$

$p(\mathbf{y}_t) = p(\mathbf{W}\mathbf{x}_t)$ and $Q_M = \prod_1^M p(y_{mt}) = \prod_1^M p(\mathbf{w}_m\mathbf{x}_t)$. Thus, the CCS-DIV as a contrast function, with the built-in convexity parameter α, is

$$D_{CCS}(P_J, Q_M, \alpha)$$
$$= \log \frac{\iint f^2(P_J) dy_1 \ldots dy_M \cdot \iint f^2(Q_M) dy_1 \ldots dy_M}{[\iint f(P_J) \cdot f(Q_M) dy_1 \ldots dy_M]^2}$$
$$= \log \frac{\iint f^2(p(\mathbf{W}\mathbf{x}_t)) dy_1 \ldots dy_M \cdot \iint f^2(\prod_1^M p(y_{mt})) dy_1 \ldots dy_M}{[\iint f(p(\mathbf{W}\mathbf{x}_t)) \cdot f(\prod_1^M p(y_{mt})) dy_1 \ldots dy_M]^2}$$

By using the Lebesgue measure to approximate the integral with respect to the joint distribution of $\mathbf{y}_t = \{y_1, y_2, \ldots, y_M\}$, the contrast function becomes

$$D_{CCS}(P_J, Q_M, \alpha)$$
$$= \log \frac{\sum_1^T f^2(p(\mathbf{W}\mathbf{x}_t)) \cdot \sum_1^T f^2(\prod_1^M (p(\mathbf{w}_m\mathbf{x}_t)))}{[\sum_1^T f(p(\mathbf{W}\mathbf{x}_t)) \cdot f(\prod_1^M (p(\mathbf{w}_m\mathbf{x}_t)))]^2}$$

For simplicity, let us define

$$V_1 = \sum_{t=1}^T f^2(P_J), \quad V_1' = \sum_{t=1}^T 2f(P_J)f'(P_J)P_J'$$
$$V_2 = \sum_{t=1}^T f^2(Q_M), \quad V_2' = \sum_{t=1}^T 2f(Q_M)f'(Q_M)Q_M'$$
$$V_3 = \sum_{t=1}^T f(P_J)f(Q_M),$$
$$V_3' = \sum_{t=1}^T f'(P_J)f(Q_M)P_J' + \sum_{t=1}^T f(P_J)f'(Q_M)Q_M'$$

and the convex function, e.g., is

$$f(t) = \frac{4}{1-\alpha^2}\left[\frac{1-\alpha}{2} + \frac{1+\alpha}{2}t - t^{\frac{1+\alpha}{2}}\right]$$
$$f'(t) = \frac{2}{1-\alpha}\left[1 - t^{\alpha-1/2}\right]$$

then,

$$P_J = p(\mathbf{W}\mathbf{x}_t) \text{ and } Q_M = \prod_{m=1}^M p(\mathbf{w}_m\mathbf{x}_t)$$

$$P_J' = \frac{\partial P_J}{\partial w_{ml}} = -\frac{p(\mathbf{x}_t)}{|\det(\mathbf{W})|^2} \cdot \frac{\partial \det(\mathbf{W})}{\partial w_{ml}} \cdot \text{sign}(\det(\mathbf{W}),$$

where $\frac{\partial \det(\mathbf{W})}{\partial w_{ml}} = W_{ml}$.

*Table IV: Kurtosis Values of the different probability density functions used in the ICA comparative study*

| Signals' Notation | Kurtosis | Signals' Notation | Kurtosis |
|---|---|---|---|
| $s_1$ | −1.2116 | $s_{12}$ | −0.65419 |
| $s_2$ | 2.9324 | $s_{13}$ | −0.33421 |
| $s_3$ | −1.3995 | $s_{14}$ | −1.6935 |
| $s_4$ | 136.0108 | $s_{15}$ | −0.86239 |
| $s_5$ | 11.6452 | $s_{16}$ | −0.60566 |
| $s_6$ | 4.219 | $s_{17}$ | −0.75488 |
| $s_7$ | −1.2065 | $s_{18}$ | −0.65645 |
| $s_8$ | 3.1965 | $s_{19}$ | −0.81022 |
| $s_9$ | 3.4302 | $s_{20}$ | −0.7692 |
| $s_{10}$ | −1.3049 | $s_{21}$ | −0.27737 |
| $s_{11}$ | −1.6805 | $s_{22}$ | −0.56816 |

$$Q_M' = \frac{\partial Q_M}{\partial w_{ml}} = \left[\prod_{j=m}^M p(\mathbf{w}_j\mathbf{x}_t)\right]\frac{\partial p(\mathbf{w}_n\mathbf{x}_t)}{\partial (\mathbf{w}_n\mathbf{x}_t)} \cdot x_l.$$

where $x_l$ denotes the *lth* entry of $\mathbf{x}_t$.
Thus, we re-write the CCS-DIV as

$$D_{CCS}(P_J, Q_M, \alpha) = \log \frac{V_1 \cdot V_2}{[V_3]^2} = \log V_1 + \log V_2 - 2\log V_3$$

and its derivative becomes

$$\frac{\partial D_{CCS}(P_J, Q_M, \alpha)}{\partial w_{ml}} = \frac{V_1'}{V_1} + \frac{V_2'}{V_2} - 2 * \frac{V_3'}{V_3}$$

**Table V:** *The performance of the ICA algorithm based on the proposed divergence and other widely used ICA algorithms in terms of Amari error [2] (multiplied by 100). Each entry averages over the corresponding number of trials.*

| Dimensions | Samples | Trials | JADE | FastICA | RapidICA | RobustICA | CS | CDIV | KLDIV | CCS2 | CCS3 |
|---|---|---|---|---|---|---|---|---|---|---|---|
| 2 | 1000 | 512 | 5.6 | 7.3 | 6.1 | 7.2 | 2.5 | 2.2 | 2.3 | 2.1 | 2 |
| | 2000 | 512 | 5.1 | 5.9 | 5.5 | 6 | 1.9 | 1.7 | 1.9 | 1.8 | 1.8 |
| | 4000 | 512 | 3.1 | 4.1 | 3.5 | 4.3 | 1.7 | 1.6 | 1.5 | 1.6 | 1.4 |
| | 8000 | 512 | 2.4 | 2.6 | 2.5 | 2.6 | 1.3 | 1.2 | 1.1 | 1.4 | 1.1 |
| 4 | 1000 | 200 | 8 | 9.7 | 9.1 | 9.8 | 3.0 | 2.4 | 3.1 | 3.1 | 2.5 |
| | 2000 | 200 | 5.4 | 7.3 | 6.5 | 7.2 | 2.4 | 2.2 | 2.1 | 2.5 | 1.8 |
| | 4000 | 200 | 4.2 | 4.2 | 4.1 | 4.3 | 1.7 | 1.4 | 1.4 | 1.4 | 1.6 |
| | 8000 | 200 | 2.1 | 2.7 | 2.5 | 2.7 | 1.4 | 1.2 | 1.3 | 1.4 | 1.2 |
| 8 | 1000 | 75 | 10.5 | 10.3 | 9.6 | 11.2 | 4.6 | 3.6 | 4.2 | 4.4 | 3.2 |
| | 2000 | 75 | 8.1 | 8.0 | 7.6 | 8.2 | 3.5 | 3.1 | 3.3 | 3.2 | 3 |
| | 4000 | 75 | 5.7 | 4.1 | 4.4 | 4.9 | 2.5 | 2.3 | 2.7 | 2.6 | 2.8 |
| | 8000 | 75 | 2.7 | 3.1 | 3.0 | 3.2 | 2.3 | 2.1 | 2 | 2.1 | 1.9 |
| 16 | 1000 | 15 | 8 | 9.7 | 9.1 | 9.8 | 6.7 | 6 | 6.7 | 7.3 | 5.5 |
| | 2000 | 15 | 5.4 | 7.3 | 6.5 | 7.2 | 6.1 | 5.2 | 6 | 6.9 | 5.1 |
| | 4000 | 15 | 4.2 | 4.2 | 4.1 | 4.3 | 5.4 | 4.4 | 5.1 | 5.6 | 4.2 |
| | 8000 | 15 | 2.1 | 2.7 | 2.5 | 2.7 | 3.6 | 2.6 | 3.1 | 3.8 | 2.9 |
| 20 | 1000 | 5 | 22.3 | 21.1 | 20.1 | 26.2 | 14.1 | 9.1 | 10.1 | 13.1 | 8.9 |
| | 2000 | 5 | 15.7 | 15.6 | 15.2 | 16.2 | 7.7 | 6.7 | 7.3 | 8.3 | 7.2 |
| | 4000 | 5 | 7.8 | 7.2 | 7.1 | 7.2 | 6.2 | 7.6 | 6.4 | 6.7 | 5.3 |
| | 8000 | 5 | 4.5 | 4.1 | 3.9 | 4.0 | 2.7 | 2.2 | 2.6 | 4.4 | 2.3 |

**Table VI:** *The performance of the ICA algorithm based on the proposed divergence in terms of the Amari error (multiplied by 100). Each entry averages over the corresponding number of trials.*

| Dimensions M | Samples T | Trials | CCS3 at 0.1T | CCS3 At 0.01T | CCS3 At 0.001T | CCS3 At 1 |
|---|---|---|---|---|---|---|
| 2 | 1000 | 1024 | 4.6 | 2.9 | 2.1 | 2 |
| | 2000 | 1024 | 3.6 | 2.3 | 1.9 | 1.8 |
| | 4000 | 1024 | 2.8 | 1.9 | 1.6 | 1.4 |
| | 8000 | 1024 | 2.2 | 1.6 | 1.1 | 1.2 |
| 4 | 1000 | 250 | 5.8 | 3.8 | 2.4 | 2.5 |
| | 2000 | 250 | 5 | 2.9 | 2 | 1.8 |
| | 4000 | 250 | 3.5 | 2.5 | 1.6 | 1.6 |
| | 8000 | 250 | 2.7 | 2.2 | 1.3 | 1.3 |
| 8 | 1000 | 100 | 5.6 | 3.8 | 2.5 | 3.2 |
| | 2000 | 100 | 3.7 | 3.1 | 2.2 | 3 |
| | 4000 | 100 | 3.1 | 2.6 | 2.2 | 2.8 |
| | 8000 | 100 | 3.0 | 2.2 | 1.9 | 1.9 |
| 16 | 1000 | 25 | 20.5 | 15.8 | 8.6 | 5.5 |
| | 2000 | 25 | 12.6 | 10.1 | 7 | 5.1 |
| | 4000 | 25 | 8.6 | 8 | 4.5 | 4.2 |
| | 8000 | 25 | 5.8 | 3.9 | 1.9 | 2.9 |
| 20 | 1000 | 10 | 27.7 | 15.1 | 13.7 | 8.9 |
| | 2000 | 10 | 22.8 | 11.3 | 12 | 7.2 |
| | 4000 | 10 | 15.6 | 9 | 7.2 | 5.3 |
| | 8000 | 10 | 9.8 | 6.3 | 3 | 2.3 |

**Table VII:** *The computational load, in seconds, of the ICA algorithm based on the proposed divergence and other widely used ICA algorithms, each entry averages over the corresponding number of trials.*

| Dimensions M | Samples T | Trials | CCS3 at 0.1T | CCS3 At 0.01T | CCS3 At 0.001T | CCS3 At 1 |
|---|---|---|---|---|---|---|
| 2 | 1000 | 1024 | 0.4 | 2.8 | 29.8 | 28 |
| | 2000 | 1024 | 0.5 | 4.8 | 44.8 | 96.4 |
| | 4000 | 1024 | 0.8 | 8 | 77.9 | 342.9 |
| | 8000 | 1024 | 1.5 | 10.6 | 137 | 1073 |
| 4 | 1000 | 250 | 1.8 | 24 | 218.1 | 237.9 |
| | 2000 | 250 | 4.3 | 39 | 344.8 | 630.3 |
| | 4000 | 250 | 5.9 | 47.9 | 593.4 | 2348.6 |
| | 8000 | 250 | 10.2 | 83.6 | 1105 | 7737.1 |
| 8 | 1000 | 100 | 19.3 | 128.7 | 1053 | 1174 |
| | 2000 | 100 | 31.5 | 201.7 | 1743 | 3347 |
| | 4000 | 100 | 46.5 | 266.4 | 3109 | 11705 |
| | 8000 | 100 | 74.2 | 241.8 | 5534 | 42115 |
| 16 | 1000 | 25 | 170.6 | 909.5 | 6282 | 4376.2 |
| | 2000 | 25 | 242.3 | 1171 | 9320 | 17918.3 |
| | 4000 | 25 | 305.5 | 1403 | 14717 | 58894.6 |
| | 8000 | 25 | 329.9 | 2297 | 25658 | 10483.4 |
| 20 | 1000 | 10 | 339 | 1195.7 | 9605 | 11355.2 |
| | 2000 | 10 | 427.4 | 1724.2 | 14708 | 27504.8 |
| | 4000 | 10 | 607.6 | 2398.3 | 23634 | 52536.6 |
| | 8000 | 10 | 900 | 3754.5 | 42538 | 97312.1 |